\documentclass[submission,copyright,creativecommons]{eptcs}
\providecommand{\event}{FESCA 2017} 
\usepackage{breakurl}             
\usepackage{underscore}           
\usepackage{amsfonts,amsmath,amssymb} 
\usepackage{tabularx} 
\usepackage{pgfplots}
\usepackage{todonotes}
\usepackage{subcaption} 

\title{Exploiting Hierarchy in the Abstraction-Based Verification\\of Statecharts Using SMT Solvers}
\author{Bence Czip\'o$^{1}$ \qquad \'Akos Hajdu$^{1,2}$ \qquad Tam\'as T\'oth$^{1}$\thanks{\thanksrichter} \qquad Istv\'an Majzik$^{1}$
\institute{$^1$Department of Measurement and Information Systems \\
Budapest University of Technology and Economics, Budapest, Hungary}
\institute{$^2$MTA-BME Lend\"ulet Cyber-Physical Systems Research Group, Budapest, Hungary}
\email{czipobence@gmail.com, \{hajdua,totht,majzik\}@mit.bme.hu}
}

\newcommand{\thanksrichter}{Partially supported by Gedeon Richter's Talentum Foundation (Gy\"{o}mr\H{o}i \'ut 19-21, 1103 Budapest, Hungary).}

\def\titlerunning{Exploiting Hierarchy in the Abstraction-Based Verification of Statecharts Using SMT Solvers}
\def\authorrunning{B. Czip\'o, \'{A}. Hajdu, T. T\'oth, I. Majzik}
\begin{document}
\maketitle

\newcommand{\reals}{\mathbb{R}}
\newcommand{\ints}{\mathbb{Z}}
\newcommand{\posints}{\mathbb{N}}
\newcommand{\rats}{\mathbb{Q}}
\newcommand{\true}{\top}
\newcommand{\false}{\bot}
\newcommand{\andlog}{\wedge}
\newcommand{\orlog}{\vee}
\newcommand{\setsize}[1]{\left\vert{#1}\right\vert}
\newcommand{\pl}{\textsf{PL}}
\newcommand{\fol}{\textsf{FOL}}
\newcommand{\suchthat}{\: | \:}
\newcommand{\ceil}[1]{\left \lceil #1 \right \rceil}

\newcommand{\stateset}{S}
\newcommand{\fsmalphabet}[0]{\Sigma}
\newcommand{\stateinstance}{s}
\newcommand{\initialstate}{\stateinstance_0}
\newcommand{\transitionset}{\mathit{Tr}}
\newcommand{\transitioninstance}{t}
\newcommand{\transsrc}{\mathit{src}}
\newcommand{\transtrg}{\mathit{trgt}}
\newcommand{\symbolfunction}{\mathit{sym}}
\newcommand{\actionset}{\mathit{Act}}
\newcommand{\actioninstance}{a}
\newcommand{\actionfunc}{\mathit{act}}
\newcommand{\skipaction}{\mathit{skip}}
\newcommand{\inputfunction}{\mathit{input}_{\smname}}
\newcommand{\outputfunction}{O}
\newcommand{\smname}{M}
\newcommand{\smtuple}{\smname = (\stateset,  \fsmalphabet, \transitionset, \initialstate)}
\newcommand{\mealysmname}{\smname_{\mathit{Mealy}}}
\newcommand{\mealytuple}{\mealysmname = (\stateset, \fsmalphabet, \transitionset, \initialstate, \actionset)}
\newcommand{\mooresmname}{\smname_{\mathit{Moore}}}
\newcommand{\mooretuple}{\mooresmname = (\stateset, \fsmalphabet, \transitionset, \outputfunction, \initialstate, \actionset)}
\newcommand{\morestates}{\omega}
\newcommand{\regionset}{R}
\newcommand{\regioninstance}{r}
\newcommand{\hierfunc}{\mathit{par}}
\newcommand{\invhier}{\mathit{chld}}
\newcommand{\hierroot}{\mathit{root}}
\newcommand{\eventset}{\mathit{EV}}
\newcommand{\eventinstance}{e}
\newcommand{\trigfunc}{\mathit{trig}}
\newcommand{\defaultevent}{\epsilon}
\newcommand{\guardset}{G}
\newcommand{\guardinstance}{g}
\newcommand{\guardfunc}{\mathit{grd}}
\newcommand{\initstates}{I} 
\newcommand{\statechartvars}{V}
\newcommand{\variableinstance}{v}
\newcommand{\statechartname}{\mathit{Sc}}
\newcommand{\statecharttuple}{\statechartname=(\stateset, \regionset, \hierfunc, \initstates, \statechartvars, \transitionset)}
\newcommand{\statecharttuplePrime}{\statechartname'=(\stateset', \regionset', \hierfunc', \initstates', \statechartvars', \transitionset')}
\newcommand{\statecharttupleHat}{\hat{\statechartname}=(\hat{\stateset}, \hat{\regionset}, \hat{\hierfunc}, \hat{\initstates}, \hat{\statechartvars}, \hat{\transitionset})}
\newcommand{\evraise}{\mathit{raise}}

\newcommand{\activestates}{\morestates}
\newcommand{\sconfname}{c}
\newcommand{\sconfsetname}{C}
\newcommand{\sconfset}{{C_{\statechartname}}}
\newcommand{\sconfreach}{{C_R}}
\newcommand{\sconfsetp}[1]{C_{#1}}
\newcommand{\activeevents}{\rho}
\newcommand{\valuefunction}{\alpha}
\newcommand{\scvarvalue}{\mathit{Val}}
\newcommand{\valueinstance}{\mathit{val}}
\newcommand{\valuefunctiondef}{\valuefunction: \statechartvars \mapsto \scvarvalue}
\newcommand{\sconftuple}{\sconfname=(\activestates,\activeevents, \valuefunction)}
\newcommand{\sconftupleHat}{\hat{\sconfname}=(\hat{\activestates},\hat{\activeevents}, \hat{\valuefunction})}
\newcommand{\sconftupleIdx}[1]{\sconfname_{#1}=(\activestates_{#1},\activeevents_{#1}, \valuefunction_{#1})}

\newcommand{\initialsconf}{\mathit{c}_{\mathit{I}}}
\newcommand{\transrel}{\mathit{N}}
\newcommand{\pathname}{\pi}
\newcommand{\pathtuple}[1]{\pathname=(\sconfname_0, \sconfname_1, \ldots, \sconfname_{#1})}
\newcommand{\pathtupleHat}[1]{\hat{\pathname}=(\hat{\sconfname}_{0}, \hat{\sconfname}_{1}, \ldots, \hat{\sconfname}_{#1})}
\newcommand{\intrprt}{\mathcal{I}}
\newcommand{\propvarset}{\mathcal{L}_0}

\newcommand \formulasym {\psi}
\newcommand \formulainstf {\formulasym_1}
\newcommand \formulainsts {\formulasym_2}

\newcommand{\folformsym}{\psi}
\newcommand{\domain}{D}
\newcommand{\assignement}{\alpha}

\newcommand{\theory}{\mathcal{T}}
\newcommand{\axioms}{\mathcal{A}}
\newcommand{\fotformsym}{\formulasym}

\newcommand{\bitvectorset}{\mathit{BV}}
\newcommand{\ibitvectorset}[1]{{\bitvectorset}_{#1}}
\newcommand{\bitvectorname}{\mathit{bv}}
\newcommand{\bitvectorbit}[1]{\bitvectorname[#1]}
\newcommand{\bvec}[1]{#1}
\newcommand{\literalfunc}{\mathit{lit}}
\newcommand{\encodefunc}{\mathit{enc}}
\newcommand{\bvtoform}{f}
\newcommand{\formfuncsym}{\mathit{form}}
\newcommand{\formfunc}[1]{\formfuncsym(#1)}

\newcommand{\elapsed}{k}

\newcommand{\isbadstate}{p}

\newcommand{\combfunc}{\mathit{comb}}

\newcommand{\dc}{X}
\newcommand{\minbits}[1]{\mathit{bits(#1)}}
\newcommand{\regionoffset}[1]{\mathit{offs}(#1)}
\newcommand{\statedepth}{\mathit{depth}}
\newcommand{\maxsttdepth}{d}
\newcommand{\leveloffset}{\mathit{offs}}
\newcommand{\encpartfunc}{\mathit{enc}_p}
\newcommand{\varfunc}{\folformsym_{\statechartvars}}

\newcommand{\ancfunc}{\mathit{anc}}
\newcommand{\descfunc}{\mathit{desc}}

\newcommand{\astatefunc}{\folformsym_{\activestates}}

\newcommand{\initfunc}{\mathit{init}}
\newcommand{\targetfunc}{\mathit{target}}

\newcommand{\abstractfunc}{\mathbf{h}}
\newcommand{\invabsfunc}{{\abstractfunc}^{-1}}
\newcommand{\absvar}{generic abstraction}
\newcommand{\absstate}{states-only abstraction}
\newcommand{\unbound}{\mathit{unb}}
\newcommand{\bound}{\mathit{bound}}

\newcommand{\classname}[1]{\texttt{#1}}
\newcommand{\method}[1]{\texttt{#1}}
\newcommand{\inline}[1]{\code{#1}}
\newcommand{\thetatool}{\textsf{theta}}
\newcommand{\manyatonceNonPop}{MON}
\newcommand{\manyatoncePop}{MOP}
\newcommand{\oneatonce}{OAO}
\newcommand{\bmc}{BMC}
\newcommand{\hieronly}{STT}
\newcommand{\hierfirst}{GEN}
\begin{abstract}
Statecharts are frequently used as a modeling formalism in the design of state-based systems. Formal verification techniques are also often applied to prove certain properties about the behavior of the system. One of the most efficient techniques for formal verification is Counterexample-Guided Abstraction Refinement (CEGAR), which reduces the complexity of systems by automatically building and refining abstractions. In our paper we present a novel adaptation of the CEGAR approach to hierarchical statechart models. First we introduce an encoding of the statechart to logical formulas that preserves information about the state hierarchy. Based on this encoding we propose abstraction and refinement techniques that utilize the hierarchical structure of statecharts and also handle variables in the model. The encoding allows us to use SMT solvers for the systematic exploration and verification of the abstract model, including also bounded model checking. We demonstrate the applicability and efficiency of our abstraction techniques with measurements on an industry-motivated example.
\end{abstract}
\section{Introduction}
Statecharts are frequently used for modeling and designing state-based systems. Such systems also appear in safety critical domains, thus ensuring their correct operation is gaining increasing importance. Formal verification techniques (such as model checking) can yield mathematically precise proofs regarding the correctness of a model of the system. A widely used requirement is safety, where the purpose of verification is to check if a given erroneous state configuration is reachable during the operation of a system. However, a typical drawback of using formal verification techniques is their high computational complexity, as the set of possible configurations for a system can be unmanageably large or even infinite. A possible solution to overcome this issue is to use abstraction, which is a generic technique for reducing complexity by hiding details that are not relevant for the property to be verified. However, it is a difficult task to find the proper precision of abstraction, which shall be coarse enough to avoid complexity issues but fine enough to prove the desired property. Counterexample-Guided Abstraction Refinement (CEGAR) is an automatic technique that initially starts with a coarse abstraction and refines it iteratively based on the counterexamples until the proper precision is obtained~\cite{clarke03}. CEGAR was first described for transition systems \cite{clarke03} but since then it has been applied in various fields of verification~\cite{beyer13, clarke04, hajdu15}.

In our paper we present a novel adaptation of the CEGAR approach to the reachability analysis of hierarchical statecharts. We first define an encoding of the statechart to logical formulas that preserves information about the hierarchy, such as parallel regions and composite states. Our approach also supports some additional elements of statecharts, including variables, events, guards and actions. Based on this encoding we propose an abstraction over the hierarchical structure by only expanding composite states until a given depth. Refinement is performed by increasing the depth for certain states along a spurious counterexample. Furthermore, we also combine our state-based abstraction technique with the variable abstraction of Clarke et al.~\cite{clarke04} to efficiently handle variables in the abstract statechart. The main novelty of our approach is that the encoding allows us to use SMT solvers~\cite{bradley} for the systematic exploration and bounded model checking of the abstract state space. We evaluate and demonstrate the applicability and scalability of our algorithms by performing reachability queries on an industry-motivated example.

The rest of this section discusses related work. Section~\ref{sect:back} introduces preliminaries of our work. Section~\ref{sect:encoding} presents our encoding of statecharts to logical formulas. Section~\ref{sect:cegar} describes the adaptation of CEGAR to statecharts. Section~\ref{sect:eval} evaluates the algorithms and Section~\ref{sect:conclusions} concludes our work.

\paragraph{Related work.}
Several works in the literature address the formal verification of statecharts. Based on a survey~\cite{purandar2004} most approaches flatten the hierarchy of the statechart or transform the problem to the input language of a model checker such as SMV~\cite{chan1998} or SPIN~\cite{latella1999}. The disadvantage of these approaches is that the information in the state hierarchy is not preserved and it is often difficult to interpret the results on the original statechart. Alur et al.~\cite{alur02} exploit hierarchy, but they work with hierarchical reactive modules, where hierarchy has a bit different semantics than in statecharts: submodules can interact through interfaces and concurrency is only allowed at the top level.

The work of Meller et al.~\cite{meller2014, meller16} is the most related to our current paper. They also defined a CEGAR-like approach for statecharts, supporting a wide range of their elements. They focus on LTL$_x$ model checking, while our approach currently only targets reachability. They use a model-to-model transformation, which means that the abstraction of a statechart is also a statechart similarly to our approach. They abstract a composite state using its interface, whereas we treat abstracted composite states as a simple state. The main difference however, is that in their approach the abstract model is transformed to the input language of a model checker, whereas in our method we encode the abstract statechart as SMT formulas, allowing us to use SMT solvers to perform CEGAR and to utilize the power of SMT-based model checking.

The recent work of Helke and Kamm\"{u}ller~\cite{helke2016} also defines abstractions over statecharts for the universal fragment of CTL model checking, which is more general than our reachability analysis. However, their main focus is on only abstracting the data and preserving the structure of the statechart, whereas in our approach abstraction on the structure is also a key feature.

\section{Background}
\label{sect:back}

In this section we first present hierarchical statecharts as the formalism used in our work (Section~\ref{sect:back:sc}). Then, we describe model checking (Section~\ref{sect:back:modelcheck}) and we introduce Counterexample-Guided Abstraction Refinement (Section~\ref{sect:back:CEGAR}), an efficient model checking technique. 

\subsection{Hierarchical Statecharts}
\label{sect:back:sc}
In our work we describe state-based event-driven behavior of systems using \emph{hierarchical statecharts}~\cite{meller16}. Expressions and variables of the statechart are based on first order logic (FOL) \cite{bradley}. Let $\fol$ denote the set of all first order logic formulas. Let the formula $\folformsym \in \fol$ be a first order formula and $V = \{ \variableinstance_{1}, \variableinstance_{2},  \ldots, \variableinstance_{k} \}$ be the set of the variables appearing in $\folformsym$. Let $V_{i}$ represent the indexed version of the variables, i.e., $V_{i} = \{\variableinstance_{1, i}, \variableinstance_{2, i}, \ldots, \variableinstance_{k, i}\}$, and let $\folformsym_{i}$ denote the formula, where each variable $\variableinstance_{j} \in V$ is replaced by $\variableinstance_{j, i}$ from $V_{i}$. For example if $\folformsym = \variableinstance_{1} \wedge \variableinstance_{2}$ then $\folformsym_4 = \variableinstance_{1, 4} \wedge \variableinstance_{2, 4}$.

\paragraph{Hierarchical statechart.} Formally, a hierarchical statechart \cite{meller16} is a tuple $\statecharttuple$ where
	\begin{itemize}
		\item $\stateset$ is the finite set of \emph{states} in the statechart,
		\item $\regionset$ is a finite set of \emph{regions} in the statechart,
		\item $\hierfunc: (\stateset \cup \regionset) \mapsto  (\stateset \cup \regionset \cup \{\hierroot\})$ is the \emph{hierarchy function} that assigns a parent (container) region to every state, and a parent (container) state or a distinguished \emph{root} element of the statechart to every region in the statechart,
		\item $\initstates: \regionset \mapsto \stateset$ assigns an \emph{initial state} to each region in the statechart,
		\item $\statechartvars$ is the set of \emph{variables} appearing in guards and actions,
		\item $\transitionset \subseteq \stateset \times \stateset \times \eventset \times \guardset \times \actionset$ is the set of \emph{transitions}, where $\eventset$ is the set of \emph{events}, $\guardset \subseteq \fol$ is the set of \emph{guard} expressions and $\actionset$ is the set of \emph{actions}.
	\end{itemize}
For a transition $\transitioninstance = (\stateinstance, \stateinstance', \eventinstance, \guardinstance, \actioninstance) \in \transitionset$, let $\transsrc(\transitioninstance) = \stateinstance$ and $\transtrg(\transitioninstance) = \stateinstance'$ denote its \emph{source} and \emph{target} states, let $\trigfunc(\transitioninstance) = \eventinstance$ denote its \emph{trigger} event, let $\guardfunc(\transitioninstance) = \guardinstance$ denote its guard expression, and let $\actionfunc(\transitioninstance) = \actioninstance$ denote the action executed when $\transitioninstance$ fires. In our current work an action is either an event raising $\evraise(\eventinstance)$ (where $\eventinstance \in \eventset$) or a variable assignment $\variableinstance := \folformsym$ (where $\folformsym \in \fol)$.

\begin{figure}[htb]
	\centering
	\includegraphics[width= 12cm, keepaspectratio]{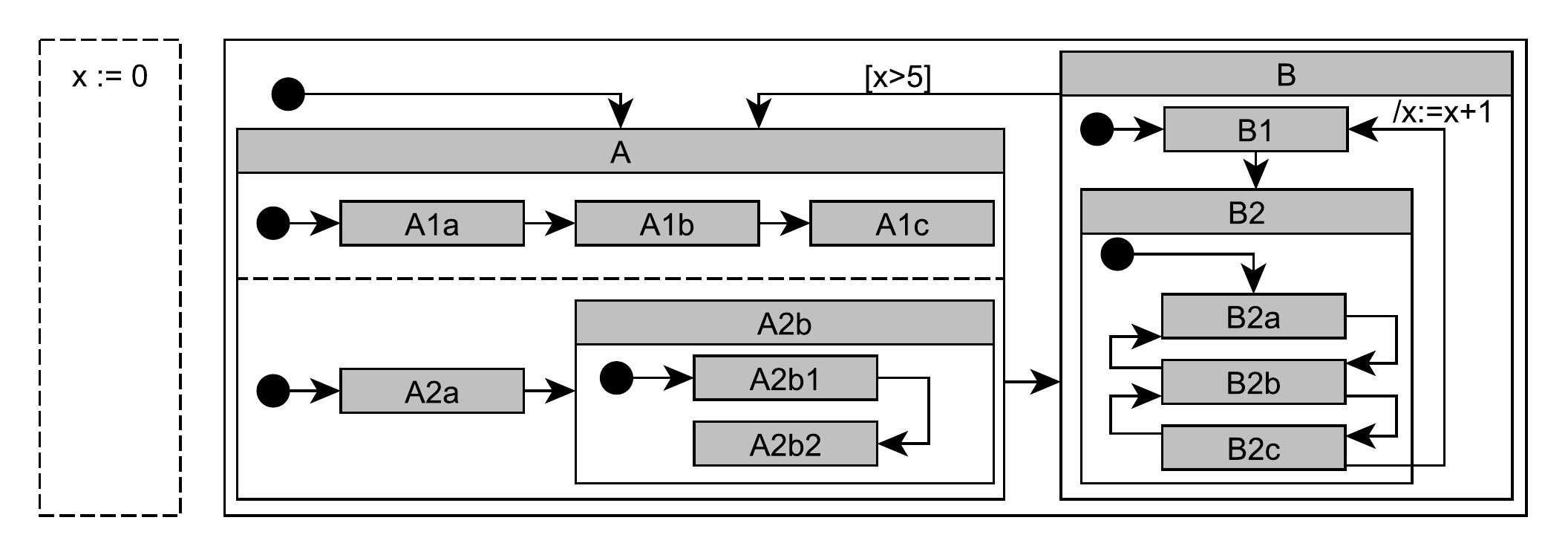}
	\caption{Example statechart with parallel regions and hierarchy.}
	\label{fig:simplExSc}
\end{figure}

An example statechart can be seen in Figure~\ref{fig:simplExSc}. In their visual representation, states are marked with rectangles and regions are marked with dashed lines, but only if there are multiple regions contained in a state, for example in case of composite state $A$. Initial states are denoted by black dots and transitions are represented by arrows. The initial values of variables are described in a dashed block on the left of the statechart ($x:=0$). The transition from state $B$ to $A$ presents an example for guard ($x > 5$) whereas the transition from state $B2c$ to $B1$ has an assignment action ($x := x + 1$).

Let $\invhier$ denote the opposite direction of the hierarchy, i.e., $\invhier(x) = \{y \suchthat \hierfunc(y) = x \}$. The set of \emph{ancestor states} of a state $\stateinstance \in \stateset$ is $\ancfunc(\stateinstance) = \{\hierfunc(\hierfunc(\stateinstance))\} \cup \ancfunc(\hierfunc(\hierfunc(\stateinstance)))$, i.e., its parent state and the ancestors of its parent. The set of \emph{descendant states} for a state $\stateinstance$ is $\descfunc(\stateinstance) = \{\stateinstance' \suchthat \stateinstance' \in \stateset, \ancfunc(\stateinstance') = \stateinstance \}$, i.e., states $\stateinstance'$ for which $s$ is an ancestor. A state $\stateinstance$ is \emph{simple} if $|\descfunc(\stateinstance)| = 0$ and \emph{composite} otherwise.

Ancestors and descendants can be defined for regions as well. The state $\stateinstance$ is an ancestor to region $\regioninstance$ ($\stateinstance \in \ancfunc(\regioninstance)$) if $\stateinstance = \hierfunc(\regioninstance)$ or $\stateinstance \in \ancfunc(\hierfunc(\regioninstance))$, and the state $\stateinstance$ is the descendant of region $\regioninstance$ ($\stateinstance \in \descfunc(\regioninstance)$), if there exists $\stateinstance' \in \invhier(\regioninstance)$ such that $\stateinstance' = \stateinstance$ or $\stateinstance' \in \ancfunc (\stateinstance)$.

The depth function $\statedepth: \stateset \mapsto \posints$ assigns the number of its ancestor states (including the root object) to a state. Inductively defined, $\statedepth(\hierroot) = 0$, and for every $\stateinstance \in \stateset$, $\statedepth(\stateinstance) = \statedepth(\hierfunc(\hierfunc(\stateinstance))) + 1$. The integer $\maxsttdepth = \max(\{ \statedepth(\stateinstance) \: | \: \stateinstance \in \stateset \})$ is the \emph{maximum depth} of the statechart.
Let the $i$-th \emph{level} of the hierarchy refer to the set of states with depth $i$, i.e., $\{\stateinstance \in \stateset \suchthat \statedepth(\stateinstance) = i \}$.
In case of the statechart in Figure~\ref{fig:simplExSc}, the depth of state $A$ is 1, $\mathit{A1a}$ is $2$ and $\mathit{A2b2}$ is $3$. The states of level $1$ are $\{A, B\}$, level $2$ are $\{A1a,A1b,A1c,A2a,A2b,B1,B2\}$ and level $3$ are $\{A2b1,A2b2,B2a,B2b,B2c\}$.

\paragraph{Configuration.} A \emph{configuration} of a statechart is a tuple $\sconftuple$ where 
\begin{itemize}
	\item $\activestates \subseteq S$ is a set of active states such that each top level region contains exactly one active state and child regions of an active state must also contain exactly one active state,
	\item $\activeevents \subseteq \eventset$ is the set of currently active events on the input of the statechart,
	\item $\valuefunction$ is the assignment to the variables $\statechartvars$.
\end{itemize}
Let $\initialsconf$ denote the initial configuration of a statechart (determined by $\initstates$), and let $\sconfset=\{\sconfname_1, \sconfname_2,\ldots\}$ denote all possible configurations of $\statechartname$. Note, that although $S$ and $R$ is finite, unbounded variables in $V$ can make $\sconfset$ infinite. Considering the example presented in Figure~\ref{fig:simplExSc}, an example configuration is $\{ \{A, \mathit{A1a}, \mathit{A2a}\}, \emptyset, \{ x = 0 \} \}$, which is also the initial configuration.

\paragraph{Transition Relation.}

The \emph{transition relation} of a statechart $\statechartname$ is a set $\transrel \subseteq \sconfset \times \sconfset$. Given two configurations $\sconftupleIdx{1}$ and $\sconftupleIdx{2}$, $(\sconfname_1, \sconfname_{2}) \in \transrel$ if a transition $\transitioninstance \in \transitionset$ exists for which the following conditions hold.
\begin{itemize}
	\item The source state $\transsrc(\transitioninstance)$ of $\transitioninstance$ is active in $\sconfname_1$ ($\transsrc(\transitioninstance) \in \activestates_1$), its guard $\guardfunc(\transitioninstance)$ evaluates to true under $\valuefunction_1$ and its trigger event is present on the input ($\trigfunc(\transitioninstance) \in \activeevents_1$),
	\item After taking $\transitioninstance$ the set of actives states $\activestates_2$ is obtained by removing its source $\transsrc(\transitioninstance)$ its ancestors $\ancfunc(\transsrc(\transitioninstance))$ and its descendants $\descfunc(\transsrc(\transitioninstance))$ from $\activestates_1$ and adding its target $\transtrg(\transitioninstance)$ and its ancestors $\ancfunc(\transtrg(\transitioninstance))$. If $\transtrg(\transitioninstance)$ is a composite state then the initial states of each of its regions are also added to $\activestates_2$ recursively. The set of active events $\activeevents_2$ is obtained by removing its trigger event $\trigfunc(\transitioninstance)$ from $\activeevents_1$ and adding the event $\eventinstance \in \eventset$ if its action $\actionfunc(\transitioninstance)$ is an event raising $\evraise(\eventinstance)$. The assignment $\valuefunction_2$ maps each variable to the same value as $\valuefunction_1$, except the variable $\variableinstance$ if the action $\actionfunc(\transitioninstance)$ is an assignment of the form $\variableinstance := \folformsym$. In this case $\variableinstance$ is mapped to $\folformsym$ in $\valuefunction_2$.
\end{itemize}

Furthermore, for a configuration $\sconfname \in \sconfset$ let $\transrel(\sconfname) = \{\sconfname' \in \sconfset \suchthat (\sconfname, \sconfname') \in \transrel\}$ denote its \emph{successors}, i.e., the set of the configurations that are reachable from $\sconfname$ with a single transition. In our work transitions do not have priority, therefore successors are selected non-deterministically.

\paragraph{Path.} A sequence of configurations $\pathtuple{n}$ is a \emph{path} in $\statechartname$ if $\sconfname_i \in \sconfset$ (for $0 \leq i \leq n$), $(\sconfname_i, \sconfname_{i+1}) \in \transrel$ (for $0 \leq i < n$) and $\sconfname_0 = \initialsconf$. The length of a path is the number of transitions occurring in the path, so the length of a path with $n+1$ configurations is $n$. For the example statechart presented in Figure~\ref{fig:simplExSc} a possible path is $\pathname=(\{ \{A,A1a,A2a\}, \emptyset, \{ x = 0 \} \}, \{ \{A, A1a, A2b, A2b1\}, \emptyset, \{ x = 0 \} \},  \{ \{B, B1\}, \emptyset, \{ x = 0 \} \})$, with a length of $2$.
A configuration $\sconfname_r$ is \emph{reachable} if there is a path $\pathtuple{n}$ leading to $\sconfname_r$, i.e, $\sconfname_n = \sconfname_r$. Let the set of all reachable configurations be denoted by $\sconfreach \subseteq \sconfset$.

\paragraph{Bit vectors.}
Bit vectors are sequences of $0$ (false) and $1$ (true) bits. The set of bit vectors of length n is denoted by $\bitvectorset_n$. The $i$th bit in the vector $\bitvectorname \in \bitvectorset_n$ is denoted by $\bitvectorbit{i}$.
Bit vectors can be extended with \emph{don't care} bits ($\dc$). Two bits are \emph{conflicting} if they are not the same and none of them is $\dc$. The combination of non-conflicting bits $b_1$ and $b_2$ is $b_1$ if $b_1 = b_2$ or $b_2 = \dc$ and $b_2$ otherwise. Two bit vectors are conflicting if they contain conflicting bit pairs at any position. If two bit vectors (of the same length) are non-conflicting, they are \emph{combinable}, and their combination is their bitwise combination. 

\subsection{Model Checking}
\label{sect:back:modelcheck}

Model checking \cite{clarke99} is a technique to verify systems against given requirements by systematically traversing the state space of the system. In our paper we focus on \emph{safety requirements}. A statechart $\statechartname$ is \emph{safe} for a predicate $p \colon \sconfset \mapsto \{\top,\bot\}$ over $\sconfset$ if for every reachable configuration $\sconfname_{r} \in \sconfreach$, $p(\sconfname_r) = \true$ holds. If the statechart is not safe, a \emph{counterexample} can be found, which is a path $\pathtuple{n}$ with $p(\sconfname_n) = \false$. Reachability and safety are opposites: a system is safe if no ``bad'' configuration is reachable.

\paragraph{State Space Exploration.}
The most basic way of checking a safety requirement is to systematically enumerate the set of reachable configurations $\sconfreach$ and to check if the predicate $p$ holds. This can be done by first starting from the initial configuration ($\sconfreach_0 = \{\initialsconf\}$) and then iteratively adding configurations that are reachable in one step from the already reached configurations ($\sconfreach_{i+1} = \sconfreach_i \cup \{\sconfname' \suchthat (\sconfname, \sconfname') \in \transrel, \sconfname \in \sconfreach_i \}$) until a fixpoint is reached ($\sconfreach_{i+1} = \sconfreach_{i}$). However, as $\sconfreach$ can be large or even infinite, this method is only applicable for large systems with additional techniques, for example CEGAR (Section~\ref{sect:back:CEGAR}).

\paragraph{Bounded Model Checking.}
\emph{Bounded Model Checking} (BMC) \cite{biere99} is an iterative algorithm to check if a safety requirement holds within a given bound $k$. A configuration $\sconfname_k$ for a statechart $\statechartname$ is considered \emph{$k$-reachable} if there is a path $\pathtuple{k}$ in $\statechartname$ leading to $\sconfname_k$ with length $k$. Bounded model checking iteratively checks the safety of $k$-reachable configurations, incrementing $k$ from $0$ to an upper bound (or until a counterexample is found).

Bounded model checking is realized by transforming $k$-reachability to a SAT or SMT formula \cite{biere99} such that $\sconfname_k$ is reachable if and only if the formula is satisfiable. In case of a \emph{transition system} that only contains states ($S$), transitions ($T \subseteq S \times S$) and initial states ($I \subseteq S$) the widely used approach is to assign a unique bit vector to each state $s \in S$. Such vectors are then transformed into formulas by assigning a boolean variable to each bit and forming a conjunction of them in the following way: if a bit is $0$ then its corresponding variable is negated. Let the formula assigned to state $s \in S$ be denoted by $\formfunc{s}$. Then, a  transition $(s, s') \in T$ at step $i$ is expressed as $\formfunc{(s, s')}_i = \formfunc{s}_i \andlog \formfunc{s'}_{i+1}$, and the whole transition relation is expressed as $\formfunc{T}_i = \bigvee_{t \in T}\formfunc{t}_i$, i.e., the disjunction of formulas assigned to transitions. Reachability in $k$ steps can be decided by solving the formula $(\bigvee_{s \in I} \formfunc{s}_0) \: \wedge \: (\bigwedge_{i=0}^{k} \formfunc{T}_{i})$, which is also called as \emph{unfolding} the transition relation $k$ times. Furthermore, a satisfying assignment to the variables also determines the bit vectors, thus states along the path can be reconstructed. Consequently, this technique can also be used in state space exploration when enumerating states reachable in one step. However, this technique cannot be applied directly to hierarchical statecharts, as their configurations can contain multiple active states and they also have additional constructs like guards and actions. In the next section we propose a generalization of this transformation approach to support hierarchical statecharts and their elements.

\subsection{CEGAR}
\label{sect:back:CEGAR}
Formal verification methods face difficulties handling large and sophisticated systems as their set of reachable configurations can be large or even infinite. \emph{Abstraction} is a general mathematical approach to simplify the model checking problem by hiding irrelevant information from the system. In this paper we restrict to \emph{existential} abstractions, that are over-approximating the original system, i.e., they might introduce additional behavior. A major issue with abstraction-based methods is to find the proper precision of abstraction that is fine enough to prove the desired requirement but coarse enough to reduce complexity. \emph{Counterexample-Guided Abstraction Refinement} (CEGAR) \cite{clarke03} is a general algorithm to automatically find the required precision of the abstraction by refining it based on counterexamples. The algorithm was first described for transition systems~\cite{clarke03} but since then it has been applied in various fields~\cite{beyer13, clarke04, hajdu15}.

CEGAR-based algorithms usually consist of four main steps. The first step is to create an \emph{initial abstraction} from the original system. Then the abstract system is \emph{checked} against the requirement (for example using state space exploration or BMC). If the requirement holds, due to the existential property of the abstraction, it also holds for the original system. Otherwise, an abstract counterexample exists, that has to be checked whether it is feasible in the original system (\emph{concretization}). If it succeeds, a counterexample is found in the  original system (witnessing that the requirement does not hold). Otherwise, the counterexample is only caused by a behavior introduced by the abstraction, so it is called \emph{spurious} and the abstraction has to be \emph{refined}. After the refinement, the abstract system can be checked again and this process is repeated.

\section{Hierarchy Preserving Encoding of Statecharts}
\label{sect:encoding}
In this section we present a novel technique to transform hierarchical statecharts to logical formulas. We first describe an encoding that assigns bit vectors to states exploiting the hierarchy (Section~\ref{sect:encoding:encoding}) and then we extend the transformation of bit vectors to logical formulas supporting the additional constructs of statecharts (Section~\ref{sect:encoding:transform}).

\subsection{Encoding States to Bit Vectors}
\label{sect:encoding:encoding}
In order to assign bit vectors to the states of a statechart using the hierarchy, two main problems have to be addressed. First Section~\ref{sect:enc:enc:par} presents an encoding for statecharts containing only parallel regions (but no hierarchy) and then Section~\ref{sect:enc:enc:hier} generalizes this encoding for any kind of hierarchical statechart.

\subsubsection{Encoding Parallel Regions}
\label{sect:enc:enc:par}
Our main idea of encoding states in parallel regions is that each region gets a fixed segment in an $n$ bit long bit vector with each segment being encoded independently. This way we can refer to a single state by omitting the other segments (filling their bits with don't care values) and we can also refer to a configuration by joining the segments of the active states in each region.

Formally, for a region $\regioninstance$ let $\minbits{\regioninstance}$ denote the minimum number of bits required to assign each state in $\regioninstance$ a unique bit vector, i.e., the length of its associated segment. If there is no hierarchy (only parallelism) then $\minbits{\regioninstance} = \ceil{\log_2(|\invhier(r)|)}$. Furthermore, given a set of regions $\regionset = \{ \regioninstance_1, \regioninstance_2, \ldots \regioninstance_k \}$ let $\regionoffset{\regioninstance_i}$ denote the offset (starting position) of the segment of each region. The offset can easily by calculated for the $i$th region by summing the size of the preceding segments: $\regionoffset{\regioninstance_i} = \sum_{j = 1}^{i - 1} \minbits{\regioninstance_j}$.

Then for a non-hierarchical statechart with parallel regions $R=\{\regioninstance_1, \regioninstance_2, \ldots, \regioninstance_k \}$ let $\encodefunc \colon \stateset \mapsto \bitvectorset_n$ assign a bit vector of length $n = \sum_{j = 1}^{k} \minbits{\regioninstance_j}$ to each state in the following way. 

\begin{enumerate}
	\item First for each region $\regioninstance_i$ let $\encodefunc_i$ assign a locally unique bit vector $\encodefunc_i(\stateinstance)$ of length $\minbits{\regioninstance_i}$ to each state $\stateinstance \in \invhier(\regioninstance_i)$ of the region. This can easily be done for example by simply numbering the states from $0$ to $|\invhier(\regioninstance_i)| - 1$ and encoding them into binary form.
	\item Then for each state $\stateinstance \in \stateset$ with $\stateinstance \in \invhier(\regioninstance_i)$ let the assigned bit vector $\encodefunc(\stateinstance)$ be defined in the following way: $\encodefunc(\stateinstance)[j] = \encodefunc_i(\stateinstance)[j - \regionoffset{\regioninstance_i}]$ if $\regionoffset{\regioninstance_i} < j \leq \regionoffset{\regioninstance_i} + \minbits{\regioninstance_i}$, and $\dc$ otherwise. In other words, the associated segment is filled with the locally unique bits while other segments are filled with don't care bits.
\end{enumerate}

The advantage of our encoding is that transitions within a region can be translated to a logical formula using the encoding of their source and target states, without affecting the other regions. Furthermore, the set of active states in a configuration can also be encoded by taking the combination of the bit vectors corresponding to each state.

\subsubsection{Encoding Hierarchy}
\label{sect:enc:enc:hier}

Our main idea of encoding hierarchically nested states is to express containment in bit vectors. Similarly to parallel regions, we do this by assigning each level of the hierarchy a fixed segment in the bit vector with each segment being encoded independently. This way we can refer to a (possibly composite) state on the $i$th level by omitting the segments after the $i$th index (filling their bits with don't care values), meaning that we do not care or know about states below the $i$th level.

Formally, let $\minbits{i}$ denote the minimum number of bits required to encode the $i$th level, which can be calculated in the following way. For a region $\regioninstance \in \regionset$, let $\minbits{\regioninstance}$ be the number of minimum bits required to encode the region, assuming that each contained state is simple, i.e., $\minbits{\regioninstance} = \ceil{\log_2(|\invhier(r)|)}$ just as before. For a composite state and the root object $\stateinstance_c \in \stateset \cup \{\hierroot\}$, let $\minbits{\stateinstance_c}$ be $\sum_{\regioninstance \in \invhier(\stateinstance_c)} \minbits{\regioninstance}$, i.e., the sum of the minimum bits required to encode each child region and for a simple state $\stateinstance_{s} \in \stateset$, let $\minbits{\stateinstance_{s}}$ be $0$ as it contains no regions.

Different states on the same hierarchy level may contain different number of descendant states, therefore requiring different number of bits for encoding. We want to be able to refer to any of the states, therefore $\minbits{i} = \max(\{ \minbits{\stateinstance} \suchthat \stateinstance \in \stateset, \statedepth(\stateinstance) = i \})$. In other words, for the $i$th level we have to take the maximum of the minimum number of bits to encode each state under that level. Note, that in the deepest level, there is no composite state (otherwise there would be another level), so $\minbits{\maxsttdepth} = 0$.

As for parallel regions, an offset $\leveloffset(i)$ can be defined for each hierarchy level that determines the starting position of its encoding in the bit vector. Again, the offset can be calculated for the $i$th level by summing the size of the preceding levels: $\leveloffset(i) = \sum_{j = 0}^{i-1} \minbits{j}$. Then states can be encoded to a bit vector of length $n = \sum_{i = 0}^{\maxsttdepth} \minbits{i}$ in the following way.

\begin{enumerate}
	\item First for each level $0 \leq i \leq d$ let $\encodefunc_i$ assign a locally unique bit vector $\encodefunc_i(\stateinstance)$ of length $\minbits{i}$ to each state $\stateinstance$ on the $i$th level ($\statedepth(\stateinstance) = i$) as if they were simple states. If there are no parallel regions, this can be done by numbering, otherwise the encoding of parallel regions (presented in Section~\ref{sect:enc:enc:par}) can be applied. By convention, we always give the number $0$ to the initial states. We will rely on this convention later when transforming transitions.
	\item Then for each state $\stateinstance_i \in \stateset$ with $\statedepth(\stateinstance) = i$ and ancestors $\{\stateinstance_1, \stateinstance_2, \ldots, \stateinstance_{i-1} \}$ let the assigned bit vector $\encodefunc(\stateinstance)$ be defined as the concatenation of $\encodefunc_j(\stateinstance_j)$ for the first $\leveloffset(i) + \minbits{i}$ bits ($0 \leq j < i$) and let the remaining bits be filled with $\dc$ bits.
\end{enumerate}

The advantage of our encoding is that if the source or target of the transition is a composite state, it implicitly implies that any descendant state can also take the transition. Furthermore, the set of active states in a configuration can be encoded by taking the combination of the bit vectors corresponding to each state, since a child state has the same prefix as its parent.

\begin{figure}[htb]
	\centering
	\includegraphics[width= 12cm, keepaspectratio]{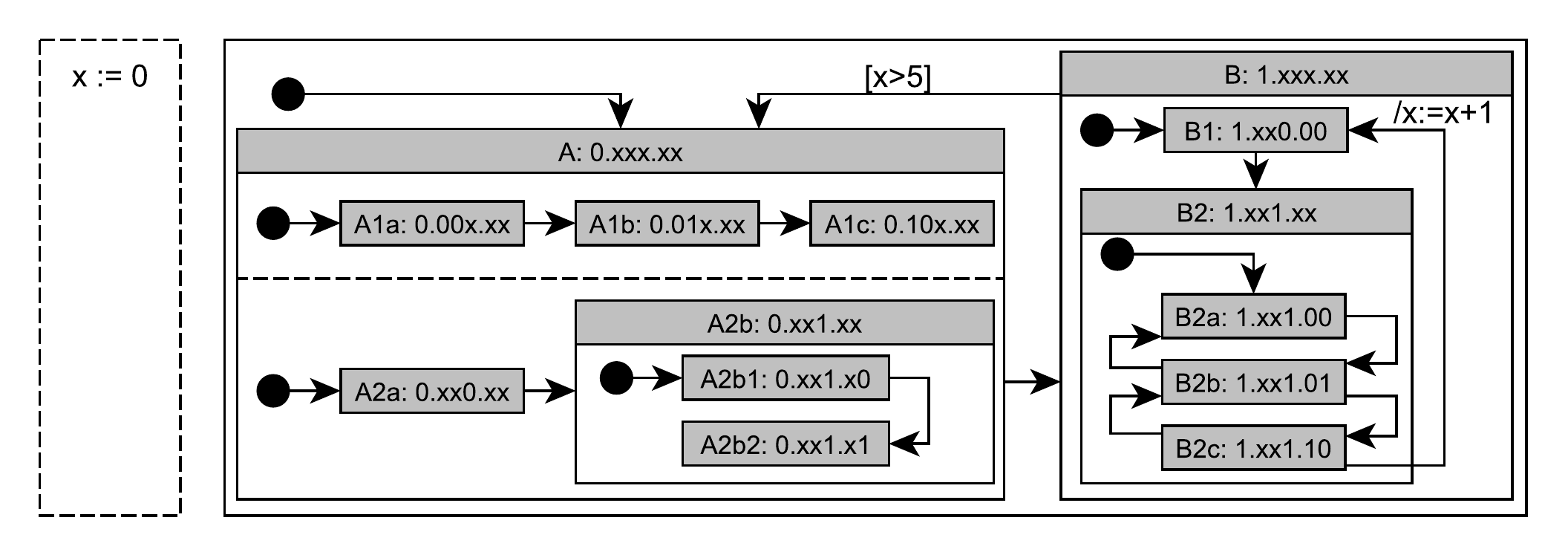}
	\caption{Encoding of an example statechart.}
	\label{fig:encEx}
\end{figure}

As an example, consider the statechart in Figure~\ref{fig:simplExSc}. A possible encoding for this statechart is presented in Figure~\ref{fig:encEx}. For the ease of understanding, segments corresponding to levels are separated by dots. On the first level, there are two states $A$ and $B$ which requires a single bit. As $A$ is the initial state, it is assigned $0$ and $B$ is assigned $1$. The second level is more complicated. $B$ only contains $B1$ and $B2$ (requiring a single bit), but $A$ has two regions containing $3$ and $2$ states respectively, which requires $2 + 1 = 3$ bits. Therefore, the second level is encoded in $\max(1,3) = 3$ bits. As $B1$ is the initial state, it gets $\dc\dc0$ and $B2$ gets $\dc\dc1$. The reason behind don't care bits here is that we only have 2 states out of the 8 that could be encoded with 3 bits. It is just a convention, replacing these $\dc$ bits with $0$ bits would not make any difference. Encoding states in $A$ require the rule for parallel regions. In the $3$ bit long segment of the second level, the first two bits are used for the top region ($A1a,A1b,A1c$) and the third bit is used for the bottom region ($A2a,A2b$). On the third level there are $2$ states in $A2b$ and $3$ states in $B2$, requiring $\max(1,2) = 2$ bits for local encoding.

As the example shows, bit vectors assigned to active states in a configuration can be combined, for example for the active states $\{A, A1c, A2b, A2b2\}$ the bit vectors assigned to the states are $\bvec{0.\dc\dc\dc.\dc\dc}$, $\bvec{0.10\dc.\dc\dc}$, $\bvec{0.\dc\dc1.\dc\dc}$, $\bvec{0.\dc\dc1.\dc1}$ respectively so their combination is $\bvec{0.101.\dc1}$.

\subsection{Transformation to Logical Formulas}
\label{sect:encoding:transform}

Based on the encoding function $\encodefunc$ defined in the previous section we now describe the transformation of statecharts to logical formulas. Let $\bvtoform \colon \bitvectorset_n \mapsto \fol$ be a function that assigns formulas to bit vectors over the set of variables $V = \{ v_1, v_2, \ldots v_n \}$ (distinct from the variables appearing in the statechart) in the following way: $\bvtoform(\bitvectorname) = \bigwedge\limits_{i=1}^n \literalfunc(\bitvectorbit{i}, i)$ where $\literalfunc(b,i)$ is $\neg v_i$ if $b=0$, $v_i$ if $b = 1$, and $\true$ if $b = \dc$. For example, the formula assigned to the bit vector $\bvec{01\dc0}$ is $\neg{v_1} \andlog v_2 \andlog \true \andlog \neg{v_4}$. Note, that due to the semantics of the $\dc$ bit, this formula has two satisfying assignments corresponding to $\bvec{0100}$ and $\bvec{0110}$.

Using the functions $\encodefunc$ and $\bvtoform$, the function $\formfuncsym$ defined in Section~\ref{sect:back:modelcheck} can be extended to hierarchical statecharts. For a state $\stateinstance$ let $\formfunc{\stateinstance}$ be $\bvtoform(\encodefunc(s))$, and for a set of active states $\activestates$ let $\formfunc{\activestates}$ be $\bvtoform(\encodefunc(\activestates))$.

To assign formulas to transitions, not only source and target states, but also trigger events, guards and actions need a subformula to be assigned. As there are finite number of events, they can also be assigned bit vectors and formulas $\formfunc{\eventinstance}$ ranging over a set of (fresh) variables. Guards are $\fol$ formulas, so in case of a guard $\guardinstance \in \guardset$, $\formfunc{\guardinstance} = \guardinstance$. In our current work we restrict the set of actions to variable assignments and event raising. An assignment $v_j := \folformsym$ in the $i$th step can be expressed with the formula $v_{j,i+1} = \folformsym_i$, whereas raising event $\evraise(\eventinstance)$ can be expressed as $\formfunc{\eventinstance}_{i+1}$. For a transition $\transitioninstance$ to fire at step $i$, each of these formulas have to be satisfied, so 

$$\formfunc{\transitioninstance}_i = \formfunc{\transsrc(\transitioninstance)}_i \andlog \formfunc{\transtrg(\transitioninstance)}_{i+1} \andlog \formfunc{\trigfunc(\transitioninstance)}_i \andlog \guardfunc(\transitioninstance)_i \andlog \formfunc{\actionfunc(\transitioninstance)}_i.$$

The formula above works well for transitions with a composite source: any descendant state will be able to take the transition. However, it is not suitable if the target of the transition is composite. In this case the transition can lead to any descendant state, whereas the semantics of statecharts specifies that the transition should lead to the initial states of each region. This problem can be easily solved by replacing $\dc$ bits with $0$ bits in $\encodefunc(\transtrg(t))$, assuming that initial states are numbered with $0$.

After defining $\formfuncsym$ for transitions, it can be defined for the whole transition relation: $\formfunc{\transitionset}_i = \bigvee_{\transitioninstance \in \transitionset}\formfunc{\transitioninstance}_i$. Furthermore, for a statechart $\statechartname$ with the initial configuration $\initialsconf$, the formula for $k$-reachability $\formfunc{\statechartname, k}$ can be defined as $\formfunc{\initialsconf} \andlog \left( \bigwedge_{i=0}^k \formfunc{\transitionset}_{i} \right)$. Note, that this formula is also suitable for state space exploration by replacing $\initialsconf$ with the actual configuration and setting $k=1$.
\section{Applying CEGAR to Hierarchical Statecharts}
\label{sect:cegar}

The techniques presented in Section~\ref{sect:encoding} can check statecharts against reachability requirements. However, the efficiency (or even termination) of those algorithms is not guaranteed for statecharts with huge or infinite state space. In this chapter we propose an adaption of the Counterexample-Guided Abstraction Refinement (CEGAR) method (Section~\ref{sect:back:CEGAR}) that can be applied to hierarchical statecharts.

\subsection{Abstraction of Statecharts}
\label{sect:scabstraction}

In order to apply CEGAR to statecharts, an over-approximating abstraction~\cite{clarke94} has to be defined first. The top-down design of systems involves an intuitive abstraction by first defining top level components and then expanding their inner behavior. For statecharts, this top-down design results in hierarchy, which provides a natural and intuitive abstraction possibility: we can obtain abstract statecharts by only expanding composite states up to a certain depth in the hierarchy. Our encoding described in the previous section supports this idea: different levels are encoded with disjoint sets of variables and formulas, therefore some of these sets do not need to be considered when the corresponding hierarchy level is not expanded.

Formally, we introduce a \emph{state abstraction function} $\abstractfunc_{\stateset}: \stateset \mapsto \stateset$ such that $\abstractfunc_{\stateset}(\stateinstance) \in \{\stateinstance\} \cup \ancfunc(\stateinstance)$, i.e., it maps each state to an \emph{abstract} state, which is either $\stateinstance$ itself or one of its ancestors. A state $\stateinstance$ is called \emph{abstracted} if $\abstractfunc_{\stateset}(\stateinstance) \in \ancfunc(\stateinstance)$ and \emph{refined} if $\abstractfunc_{\stateset}(\stateinstance) = \stateinstance$. If a composite state $\stateinstance$ is abstracted then $\abstractfunc_{\stateset}(\stateinstance') = \abstractfunc_{\stateset}(\stateinstance)$ must hold for all  $\stateinstance' \in \descfunc(\stateinstance)$, i.e., descendant states of a composite abstracted state are also abstracted and mapped to the same abstract state as their ancestor.

Besides abstracting states, we also apply abstraction to the variables based on the idea of Clarke et~al.~\cite{clarke04}, originally described for transitions systems. We adapted this approach to statecharts by defining abstraction and refinement methods. In our case abstraction means that variables $\statechartvars$ of the statechart are partitioned into two disjoint sets: \emph{visible} and \emph{invisible} variables. In the abstract statechart, only visible variables are considered. Formally, a \emph{variable abstraction function} $\abstractfunc_{\statechartvars}: \statechartvars \mapsto \{\true, \false\}$ is defined that assigns $\true$ to visible variables and $\false$ to invisible variables.

The state and variable abstraction functions can be combined into a single \emph{abstraction function} $\abstractfunc = \{\abstractfunc_{\stateset}, \abstractfunc_{\statechartvars} \}$, where $\abstractfunc(\stateinstance) = \abstractfunc_{\stateset}(\stateinstance)$ for a state $\stateinstance \in \stateset$ and $\abstractfunc(\variableinstance) = \abstractfunc_{\statechartvars}(\variableinstance)$ for a variable $\variableinstance \in \statechartvars$.

Let $\statecharttuple$ be a statechart and let $\abstractfunc$ be an abstraction function over $\statechartname$. The \emph{abstract statechart} $\abstractfunc(\statechartname) = \statecharttupleHat$ of $\statechartname$ corresponding to $\abstractfunc$ is defined in the following way.
\begin{itemize}
	\item $\hat{\stateset} = \{ \abstractfunc(\stateinstance) \suchthat \stateinstance \in \stateset \}$, i.e., only the abstract states are kept,
	\item $\hat{\regionset} = \{ \regioninstance \suchthat \exists \stateinstance \in \invhier(\regioninstance) $ such that $ \abstractfunc(\stateinstance) =\stateinstance   \}$, i.e., only regions containing refined states are kept,
	\item $\hat{\hierfunc} = \hierfunc \cap (\hat{\stateset} \times \hat{\regionset} \cup \hat{\regionset} \times \hat{\stateset})$, i.e., hierarchy is preserved between states and regions, 
	\item $\hat{\initstates}: \hat{\stateset} \mapsto \hat{\regionset}$ such that $\hat{\initstates}(\hat{\regioninstance}) = \initstates(\hat{\regioninstance})$ for each $\hat{\regioninstance} \in \hat{\regionset}$, i.e., the initial states of kept regions remain the same,
	\item $\hat{\statechartvars} = \{ \variableinstance \in \statechartvars \suchthat \abstractfunc(\variableinstance) = \true \}$, i.e., only visible variables are kept,
	\item $\hat{\transitionset} = \{ (\abstractfunc(\transsrc(\transitioninstance)),\abstractfunc(\transtrg(\transitioninstance)),\trigfunc(\transitioninstance),\guardfunc(\transitioninstance),\actionset(\transitioninstance)) \suchthat \transitioninstance \in \transitionset  \}$, i.e., sources and targets of transitions are mapped to abstract states. In guard and action expressions, each occurrence of an invisible variable is replaced with a unique constant, so constraints they represent are released.
\end{itemize}

For a set of states $\activestates$, abstraction is defined as $\abstractfunc(\activestates) = \{ \abstractfunc(\stateinstance) \suchthat \stateinstance \in \activestates \}$, i.e., the set of abstract states. For a configuration $\sconftuple$ of the statechart $\statecharttuple$, abstraction is defined as $\abstractfunc(\sconfname) = (\abstractfunc(\activestates), \activeevents, \{ \valuefunction(\variableinstance) \suchthat \variableinstance \in \hat{\statechartvars}\})$, i.e, the set of active states is abstracted, the active events are kept, and only visible variables are kept. Note that $\abstractfunc(\sconfname)$ is a configuration for $\abstractfunc(\statechartname)$ if $\sconfname$ is a configuration for $\statechartname$. Furthermore, for a set of configurations $\sconfset$ let $\abstractfunc(\sconfset) = \{\abstractfunc(\sconfname) \suchthat \sconfname \in \sconfset\}$, i.e., the set of abstracted configurations.
For a path $\pathtuple{n}$ let the \emph{abstract path} $\abstractfunc(\pathname)$ be $\hat{\pathname} = (\abstractfunc(\sconfname_0), \abstractfunc(\sconfname_1), \ldots, \abstractfunc(\sconfname_{n}))$, i.e., the sequence of abstract states.

\begin{figure}[!ht]
	\centering
	\begin{subfigure}[b]{.665\textwidth}
		\centering
		\includegraphics[width=1.0\linewidth]{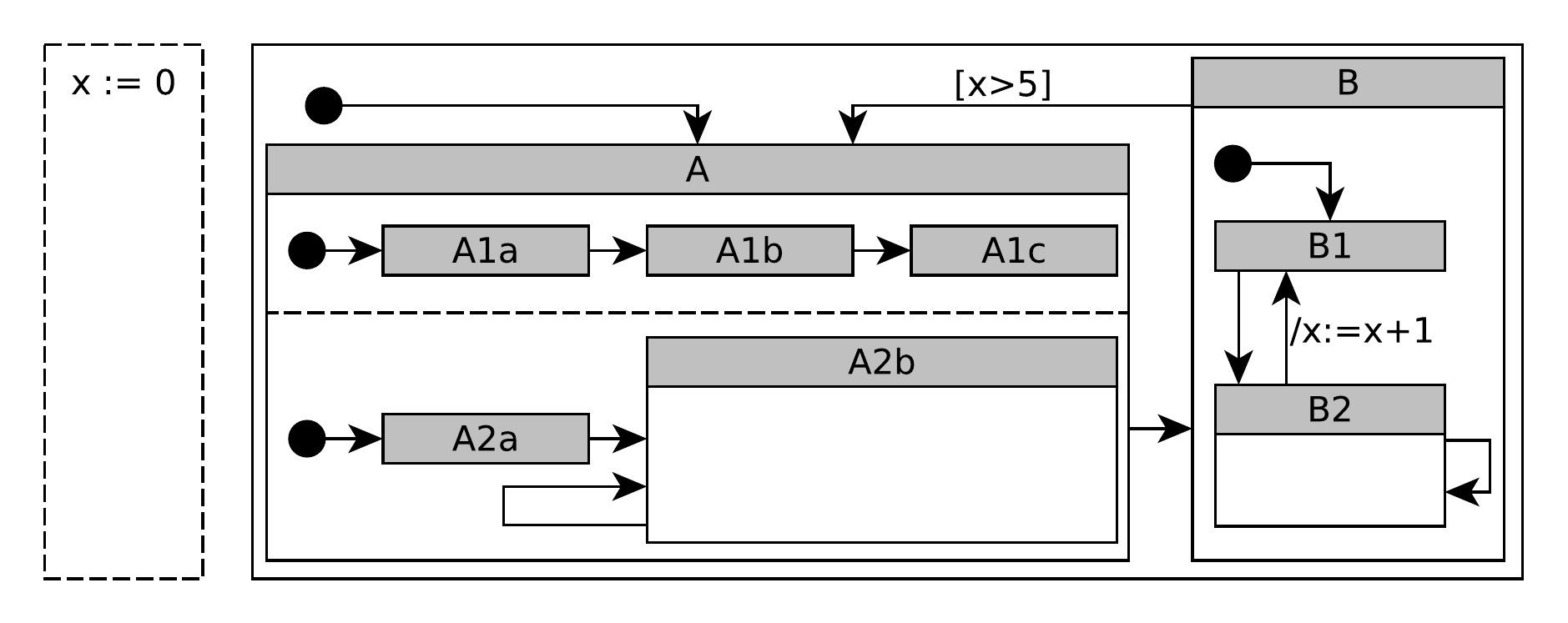}
		\caption{A finer abstraction.}
		\label{fig:exScAbsFine}
	\end{subfigure}%
	\begin{subfigure}[b]{.335\textwidth}
		\centering
		\includegraphics[width=0.7\linewidth]{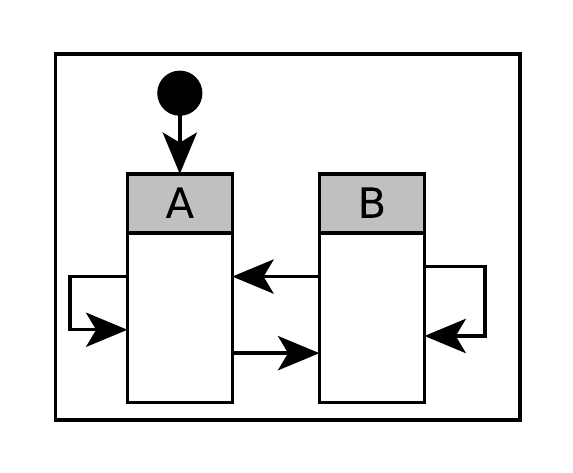}
		\caption{A coarser abstraction.}
		\label{fig:exScAbsCoarse}
	\end{subfigure}
	\caption{Two possible abstract statecharts for the example presented in Figure~\ref{fig:simplExSc}.}
	\label{fig:exScAbsColl}
\end{figure}

Recall the example statechart presented in Figure~\ref{fig:simplExSc}. Two possible abstractions for this example can be seen in Figure~\ref{fig:exScAbsColl}. The abstraction show in Figure~\ref{fig:exScAbsFine} is a finer one, with all the variables visible and states refined, except for states in $\mathit{A2b}$ and $\mathit{B2}$. In contrast, the statechart in Figure~\ref{fig:exScAbsCoarse} corresponds to a coarser abstraction, with all variables hidden, and only the top level states $A$ and $B$ being refined.

Using the abstractions defined above, we can apply the CEGAR approach to statecharts. The following sections describe the main steps of the algorithm specialized for statecharts.

\subsection{Initial Abstraction}
\label{sect:cegar:init}
The input of the algorithm is a statechart $\statecharttuple$ and a set of error configurations $\sconfsetp{f}$. This set can be explicitly given by enumerating error configurations, but it can also be defined by only bounding a subset of states and variables (e.g., $x = 1$ marks all configurations where the value of $x$ is $1$). The first step of the algorithm is to create an initial abstraction $\abstractfunc_{0} = \{{\abstractfunc_{\stateset}}_0, {\abstractfunc_{\statechartvars}}_0\}$. In our work, we defined two kind of abstractions: \emph{\absstate} that only abstracts states and \emph{\absvar} that abstracts both states and variables. Note, that the former approach can be considered as a special case for the latter one with all variables being visible ($\abstractfunc_{\statechartvars} \equiv \true$).

CEGAR-based algorithms usually start from a coarse abstraction in order to avoid complexity. For statecharts, we achieve this by abstracting each state, except states in the top-level regions. Formally, ${\abstractfunc_{\stateset}}_0(\stateinstance) = \stateinstance$ if $\statedepth(\stateinstance) = 1$ and ${\abstractfunc_{\stateset}}_0(\stateinstance) = {\abstractfunc_{\stateset}}_0(\hierfunc(\stateinstance))$ otherwise. 

The initial abstraction for the two types of abstractions is different for $\abstractfunc_{\statechartvars}$, as for \absstate{} $\abstractfunc_{\statechartvars} \equiv \true$, i.e., all variables are visible. In case of the \absvar{}, only those variables are visible that are bounded when defining the set of error configurations $\sconfsetp{f}$.

\subsection{Model Checking}
\label{sect:checking}

The input of the model checking step is the abstracted statechart $\hat{\statechartname}$, and the set of abstract error configurations $\hat{\sconfsetname}_{f}$. The verification of the model can be either performed by exploring the abstract state space or by bounded model checking as described in Section~\ref{sect:back:modelcheck}.\footnote{During bounded model checking, a limit for $k$ has to be determined, otherwise the algorithm will not terminate if the transition relation contains a cycle and the error configuration is not reachable. An upper bound for $k$ can possibly be determined by the diameter of the system~\cite{biere99}.} Due to the existential property of the abstraction, each concrete path in the statechart has its corresponding abstract path by simply mapping the states and transitions to their abstractions. Therefore, if no error configuration can be reached in the abstract statechart $\hat{\statechartname}$, then it cannot be reached in the concrete statechart and the algorithm reports that the statechart is safe. On the other hand, if an error configuration in $\hat{\sconfname} \in \hat{\sconfsetname{}}_{f}$ can be reached, an abstract path ${\hat{\pathname}}_{\hat{c}}$ leading to $\hat{\sconfname}$ is returned as a counterexample. However, it is not guaranteed that there is a corresponding concrete path $\pathtuple{n}$ in $\statechartname$ such that $\abstractfunc(\pathname) = {\hat{\pathname}}_{\hat{c}}$.

\subsection{Concretizing the Counterexample}
\label{sect:concretization}

If model checking marks an error configuration reachable, and provides an abstract counterexample $\pathtupleHat{n}$, it has to be verified if a corresponding path exists in the original statechart $\statechartname$.

We do this by iteratively checking the existence of an $0 \leq i \leq n$ long path $\pathname_i = (\sconfname_0, \sconfname_1, \ldots, \sconfname_i)$ in $\statechartname$, such that $\abstractfunc(\sconfname_j) = \hat{\sconfname}_{j}$ for $0 \leq j \leq i$. As the length of the searched path is bounded by $i$, bounded model checking can be applied here. However, the search of bounded model checking can be narrowed in this case, since we know the abstract configurations through which the concrete path must pass. Therefore, besides unfolding the transition relation, the encoding of the abstract states, events and the variable assignment formulas are also joined for each abstract configuration.

If concretization succeeds, the concretized counterexample is returned. Otherwise, if concretization succeeds until the $i$th iteration, but fails in the ($i+1$)th iteration, then $\hat{\sconfname}_{i}$ is called a \emph{failure configuration} and it is returned as a witness for the counterexample being spurious.

\subsection{Refinement}
\label{sect:cegar:refinement}

If the counterexample is spurious, the abstraction has to be refined based on the failure configuration $\sconftupleHat$. The abstraction function $\abstractfunc = \{\abstractfunc_{\stateset}, \abstractfunc_{\statechartvars} \}$ consists of two components, which can both be refined.

\begin{figure}
	\centering
	\begin{subfigure}[b]{.5\textwidth}
		\centering
		\includegraphics[width=.7\linewidth]{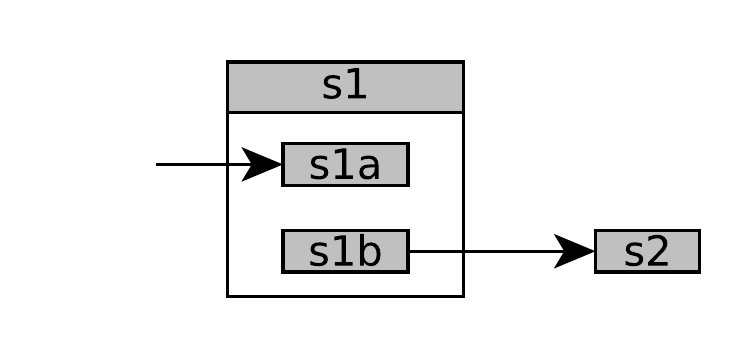}
		\caption{Failure due to state abstraction.}
		\label{fig:exRefFailStt}
	\end{subfigure}%
	\begin{subfigure}[b]{.5\textwidth}
		\centering
		\includegraphics[width=.7\linewidth]{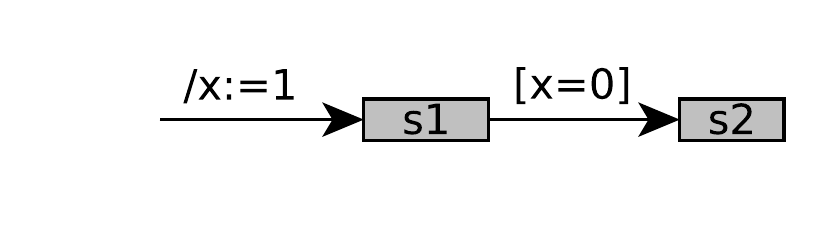}
		\caption{Failure due to variable abstraction.}
		\label{fig:exRefFailVars}
	\end{subfigure}
	\caption{Examples for failed concretization.}
	\label{fig:exRefFailCom}
\end{figure}

The state abstraction function $\abstractfunc_{\stateset}$ is refined by expanding one more level of the hierarchy in states that are included in the failure configuration, i.e., refining their direct descendants. Formally, the refined state abstraction function $\abstractfunc_{\stateset}'$ is defined as  $\abstractfunc_{\stateset}'(s) = s$ if $\hierfunc(\hierfunc(\stateinstance)) \in \hat{\activestates}$ and $\abstractfunc_{\stateset}'(s) = \abstractfunc_{\stateset}(s)$ otherwise. For the \absstate{} only this refinement technique can be used as it contains no invisible variables.
Consider the example presented in Figure~\ref{fig:exRefFailStt}. If the state $\stateinstance_1$ is not refined, the state $\stateinstance_2$ appears in reachable configurations as there is a path to it through $\stateinstance_1$, however the concretization of abstract paths through $\stateinstance_1$ will fail, as $\stateinstance_{1b}$ is not reachable and $\stateinstance_2$ is only reachable from that state.

For the \absvar{} $\abstractfunc_{\statechartvars}$ can be refined as well, however we only refine $\abstractfunc_{\statechartvars}$ if $\abstractfunc_{\stateset}$ cannot be refined anymore, i.e., we prefer to refine the hierarchy first. It can be seen that if a configuration is a failure configuration, and all the active states in the configuration are completely refined, but the execution cannot continue to the next abstract configuration, it is due to guard expressions that contain invisible variables. In this case, we refine variables that appear in guards $\guardfunc(t)$ on transitions $t$ with $\transsrc(t) \in \hat{\activestates}$, i.e., outgoing transitions from states in the error configuration.

Consider the example presented in Figure~\ref{fig:exRefFailVars}. If the variable $x$ is invisible, and from state $\stateinstance_1$ there is a transition to $\stateinstance_2$ with the guard $x=0$, but the state $\stateinstance_1$ only  appears in reachable configurations with $x=1$, then the transition cannot fire. However, if $x$ is abstracted, there is a transition from configuration $\stateinstance_1$ to $\stateinstance_2$, because there is a transition from $(\stateinstance_1,x=0)$ to $(\stateinstance_2,x=0)$.

After the refinement, execution continues with the next iteration. The CEGAR loop must eventually terminate since each refinement step either refines a state or a variable, and there are finite states and variables in the statechart. However, the model checking phase may not terminate if the abstract state space is infinite, which can happen if there is an unbounded visible variable.
\section{Evaluation}
\label{sect:eval}
A prototype of the algorithms described in Section~\ref{sect:encoding}~and~\ref{sect:cegar} has been implemented in Java using Z3~\cite{demoura08} as the underlying SMT solver. We did not compare our implementation to other tools as our current goal was to demonstrate and compare the applicability of the algorithms presented in the paper. The implementation consists of two different abstraction-refinement pairs, the \absstate{}, abbreviated in diagrams as $\hieronly$ and the \absvar{} referred to as $\hierfirst$. For model checking, four different algorithms have been implemented (summarized in Table~\ref{tab:checkmods}).

\begin{table}[ht]
	\footnotesize
	\centering
	\caption{Summary of the model checking algorithms.} \label{tab:checkmods}
	\begin{tabularx}{\linewidth}{|l|l|X|}
		\hline
		\textbf{Name} & \textbf{Abbreviation} & \textbf{Description} \\
		\hline
		Many-at-once (non-popping) & \manyatonceNonPop & A naive implementation of state space exploration, that explores all reachable configurations with an SMT solver. \\
		\hline
		Many-at-once (popping) & \manyatoncePop & A state space explorer implementation, that uses the push-pop functionality of the solver to efficiently construct the transition relation formula. \\
		\hline
		One-at-once & \oneatonce & An optimized implementation of state space exploration, that only explores one reachable configuration when turning to the solver.\\
		\hline
		Bounded model checker & \bmc & An implementation that uses bounded model checking. \\
		\hline
	\end{tabularx}
\end{table}

We tested the performance of the implemented algorithms by checking reachability queries on a statechart that represents a part of the industrial control system described in~\cite{bartha2012verification, nemeth2009}. This system represents the safety logic of a power plant, originally described as a functional block diagram. We chose this example as it contains a wide variety of the currently supported elements of statecharts: 3~hierarchy levels with a total of 27~states (5~composite, 22~simple), 9~regions (with the maximum number of parallel regions being~4), 16~variables (13~Boolean, 3~integer) and 27~transitions. Although the structure of the statechart is not so large, the parallelism and the variables induce a large number of configurations.

The metrics measured during each verification run are the time elapsed until termination (\emph{Time}), the number of CEGAR iterations (\emph{Iter}), the maximum number of explored configurations in any iteration (\emph{Confs(max)}) and the number of explored configurations at the last iteration (\emph{Confs(eve)}). Note, that the latter two metrics are not applicable for bounded model checking. An alternative metric could be the length of the path, however it only refers to the depth of the search, not the breadth.

The controller contains signal holders represented with counters, whose maximum value can be parameterized. By adjusting this parameter, the size of the state space can be varied. Measurements with different parameter values have been carried out. Table~\ref{tab:metr:211} summarizes the results with parameter value $2$.
\begin{table}[ht]
	\footnotesize
	\centering
	\caption{Results for parameter value $2$.} \label{tab:metr:211}
	\begin{tabular}{|c|c|c|c|c|c|}
		\hline
		\textbf{Checker} & \textbf{Abstraction} & \textbf{Time (s)} & \textbf{Iter}  & \textbf{Confs(max)} & \textbf{Confs(eve)} \\
		\hline
		\manyatonceNonPop & \hieronly & timeout & 2  & 8610 & 8610 \\ 
		\hline
		\manyatoncePop & \hieronly & 1398.63 & 5  & 17036 & 2855 \\ 
		\hline
		\oneatonce & \hieronly & 1250.226 & 5 & 17036 & 2855 \\  
		\hline
		\bmc & \hieronly & 211.499 & 5 &  - & - \\ 
		\hline
		\manyatonceNonPop & \hierfirst & 48.389 & 12 & 1484 & 1484 \\ 
		\hline
		\manyatoncePop & \hierfirst & 37.817 & 12 & 1484 & 1484 \\ 
		\hline
		\oneatonce & \hierfirst & 8.942 & 12 & 1484 & 1484 \\  
		\hline
		\bmc & \hierfirst & 77.478 & 12 & - & - \\ 
		\hline
	\end{tabular}
	
\end{table}

The table shows that the one-at-once state space explorer, which is based on the optimized state space exploration algorithm outperforms the other two exploration methods. Amongst the other two exploration implementations, the popping version performs slightly better than the non-popping one, which even fails to terminate with \absstate{}.
In case of the abstraction methods, the \absvar{} has better results regarding every metric with every model checker, justifying the usefulness of combining variable abstraction with state abstraction.
Note that with the \absvar{}, the bounded model checker is the least effective, however with all the variables visible, it performs the best. The improvement is relative though, as it is still approximately three times slower than with the \absvar{}. The reason behind this is that the solver can perform more efficient search in the state space than the exploring algorithms.

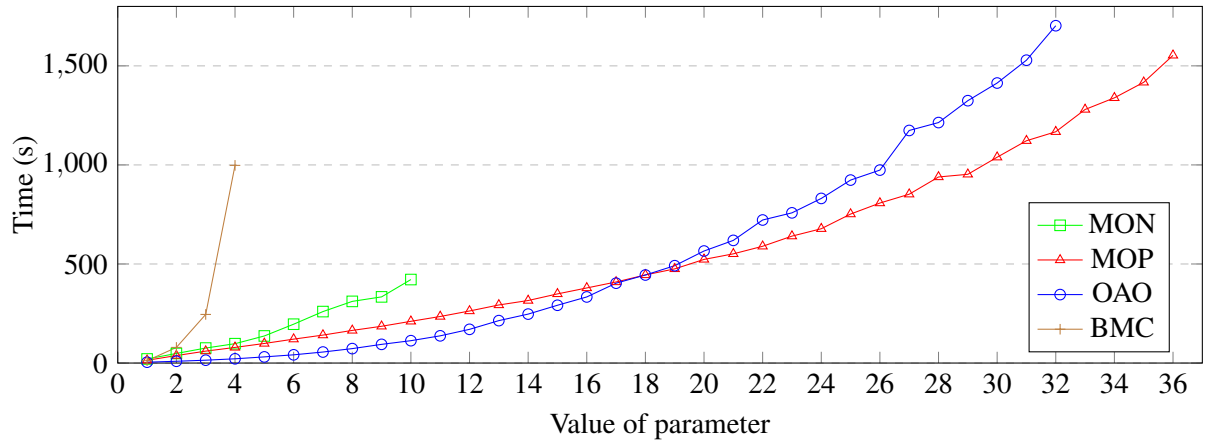
\begin{figure}[htb]
	
	\centering
	
	\begin{tikzpicture}
	\begin{axis}[
	xlabel={Value of parameter},
	ylabel={Time (s)},
	width=\textwidth,
	height=180,
	xmin=0, xmax=37,
	ymin=0, ymax=1800,
	legend pos=south east,
	ymajorgrids=true,
	grid style=dashed,
	]
	
	\addplot[
	color=green,
	mark=square,
	]
	coordinates {
		(1,20.605)
		(2,48.389)
		(3,75.696)
		(4,97.691)
		(5,136.712)
		(6,196.072)
		(7,259.800)
		(8,311.240)
		(9,333.963)
		(10,421.485)
	};
	\addlegendentry{\manyatonceNonPop}
	
	\addplot[
	color=red,
	mark=triangle,
	]
	coordinates {
		(1,13.951)
		(2,37.817)
		(3,60.110)
		(4,78.882)
		(5,98.787)
		(6,120.672)
		(7,141.130)
		(8,164.693)
		(9,185.590)
		(10,210.640)
		(11,234.943)
		(12,262.734)
		(13,293.143)
		(14,315.594)
		(15,348.995)
		(16,379.016)
		(17,409.188)
		(18,444.737)
		(19,475.228)
		(20,523.042)
		(21,551.175)
		(22,589.383)
		(23,640.607)
		(24,677.899)
		(25,751.991)
		(26,807.613)
		(27,852.162)
		(28,939.150)
		(29,952.716)
		(30,1038.889)
		(31,1122.072)
		(32,1167.370)
		(33,1279.796)
		(34,1338.570)
		(35,1417.010)
		(36,1553.082)
	};
	\addlegendentry{\manyatoncePop}

	\addplot[
	color=blue,
	mark=o,
	]
	coordinates {
		(1,4.349)
		(2,8.974)
		(3,14.382)
		(4,21.414)
		(5,31.060)
		(6,41.547)
		(7,55.620)
		(8,72.853)
		(9,94.318)
		(10,112.897)
		(11,137.370)
		(12,170.125)
		(13,214.108)
		(14,247.051)
		(15,291.653)
		(16,333.627)
		(17,403.031)
		(18,444.374)
		(19,491.406)
		(20,565.461)
		(21,619.107)
		(22,722.018)
		(23,758.093)
		(24,830.903)
		(25,923.505)
		(26,973.899)
		(27,1173.912)
		(28,1213.801)
		(29,1324.262)
		(30,1413.331)
		(31,1528.638)
		(32,1702.647)
	};
	\addlegendentry{\oneatonce}
	
	\addplot[
	color=brown,
	mark=+,
	]
	coordinates {
		(1,11.863)
		(2,78.858)
		(3,245.366)
		(4,997.769)
	};
	\addlegendentry{\bmc}
	
	\end{axis}
	\end{tikzpicture}
	\caption{Comparison of execution time for different model checking methods.}
	\label{fig:checkerComparison}
\end{figure}

In order to examine the scalability of the algorithms, they have been ran with different values of the parameter. The comparison of the execution times for the different model checking methods with \absvar{} can be seen in Figure~\ref{fig:checkerComparison}.

It turns out that the state space exploration based algorithms perform better than BMC. Amongst those three, the many-at-once implementation, that does not use the push-pop functionality of the solver is the least effective, as it is remarkably slower for every parameter value than the other two, and fails to terminate for every value greater than $10$. For small parameter values, the one-at-once explorer performs better, however for bigger parameter values, the many-at-once implementation, that uses the solver's pop-push functionality performs better.
\section{Conclusions}
\label{sect:conclusions}
In our paper we proposed a novel adaptation of the Counterexample-Guided Abstraction Refinement (CEGAR) algorithm applied to the reachability analysis of hierarchical statecharts. From the theoretical point of view, we proposed an encoding of statecharts into logical formulas that preserves information about the hierarchy of states. This encoding was effectively used in implementing abstraction and refinement techniques that utilize the hierarchical structure of statecharts. We showed that this approach allows us to use SMT solvers to check the abstract model, thus to apply full state space exploration and also bounded model checking. Furthermore, we also combined this method with the abstraction and refinement of variables in the model. On the practical side, we implemented and evaluated our new algorithms on an industry-motivated example, demonstrating their applicability.

Although the algorithms proved to be applicable, there are several opportunities for improvement. The set of the supported statechart elements can be extended with the history indicator and allowing more sophisticated event queue models. Further abstractions can be introduced, for example predicate abstraction~\cite{graf97} over variables in the statechart. The refinement methods could also be extended, for example with interpolation~\cite{mcmillan05} or unsat core-based variable refinement~\cite{leucker15}. It would also be beneficial to compare our algorithms to different approaches to see the advantages or drawbacks of CEGAR and SMT-based model checking.

\bibliographystyle{eptcs}
\bibliography{mybib}

\begin{thebibliography}{10}
\providecommand{\bibitemdeclare}[2]{}
\providecommand{\surnamestart}{}
\providecommand{\surnameend}{}
\providecommand{\urlprefix}{Available at }
\providecommand{\url}[1]{\texttt{#1}}
\providecommand{\href}[2]{\texttt{#2}}
\providecommand{\urlalt}[2]{\href{#1}{#2}}
\providecommand{\doi}[1]{doi:\urlalt{http://dx.doi.org/#1}{#1}}
\providecommand{\bibinfo}[2]{#2}

\bibitemdeclare{inbook}{alur02}
\bibitem{alur02}
\bibinfo{author}{Rajeev \surnamestart Alur\surnameend},
  \bibinfo{author}{Michael \surnamestart McDougall\surnameend} \&
  \bibinfo{author}{Zijiang \surnamestart Yang\surnameend}
  (\bibinfo{year}{2002}): \emph{\bibinfo{title}{Exploiting Behavioral Hierarchy
  for Efficient Model Checking}}, pp. \bibinfo{pages}{338--342}.
\newblock \bibinfo{series}{Lecture Notes in Computer Science},
  \bibinfo{publisher}{Springer}, \doi{10.1007/3-540-45657-0_25}.

\bibitemdeclare{inproceedings}{bartha2012verification}
\bibitem{bartha2012verification}
\bibinfo{author}{Tam\'as \surnamestart Bartha\surnameend},
  \bibinfo{author}{Andr\'as \surnamestart V\"or\"os\surnameend},
  \bibinfo{author}{Attila \surnamestart J\'ambor\surnameend} \&
  \bibinfo{author}{D\'aniel \surnamestart Darvas\surnameend}
  (\bibinfo{year}{2012}): \emph{\bibinfo{title}{Verification of an Industrial
  Safety Function Using Coloured {P}etri Nets and Model Checking}}.
\newblock In: {\sl \bibinfo{booktitle}{Proceedings of the 14th International
  Conference on Modern Information Technology in the Innovation Processes of
  the Industrial Enterprises (MITIP 2012)}}, \bibinfo{publisher}{Hungarian
  Academy of Sciences, Computer and Automation Research Institute}, pp.
  \bibinfo{pages}{472--485}.

\bibitemdeclare{incollection}{beyer13}
\bibitem{beyer13}
\bibinfo{author}{Dirk \surnamestart Beyer\surnameend} \&
  \bibinfo{author}{Stefan \surnamestart L\"{o}we\surnameend}
  (\bibinfo{year}{2013}): \emph{\bibinfo{title}{Explicit-State Software Model
  Checking Based on {CEGAR} and Interpolation}}.
\newblock In: {\sl \bibinfo{booktitle}{Fundamental Approaches to Software
  Engineering}}, {\sl \bibinfo{series}{Lecture Notes in Computer Science}}
  \bibinfo{volume}{7793}, \bibinfo{publisher}{Springer}, pp.
  \bibinfo{pages}{146--162}, \doi{10.1007/978-3-642-37057-1_11}.

\bibitemdeclare{article}{purandar2004}
\bibitem{purandar2004}
\bibinfo{author}{Purandar \surnamestart Bhaduri\surnameend} \&
  \bibinfo{author}{S.~\surnamestart Ramesh\surnameend} (\bibinfo{year}{2004}):
  \emph{\bibinfo{title}{Model Checking of Statechart Models: Survey and
  Research Directions}}.
\newblock {\sl \bibinfo{journal}{CoRR}} \bibinfo{volume}{cs.SE/0407038}.
\newblock \urlprefix\url{http://arxiv.org/abs/cs.SE/0407038}.

\bibitemdeclare{incollection}{biere99}
\bibitem{biere99}
\bibinfo{author}{Armin \surnamestart Biere\surnameend},
  \bibinfo{author}{Alessandro \surnamestart Cimatti\surnameend},
  \bibinfo{author}{Edmund \surnamestart Clarke\surnameend} \&
  \bibinfo{author}{Yunshan \surnamestart Zhu\surnameend}
  (\bibinfo{year}{1999}): \emph{\bibinfo{title}{Symbolic Model Checking without
  {BDD}s}}.
\newblock In: {\sl \bibinfo{booktitle}{Tools and Algorithms for the
  Construction and Analysis of Systems}}, {\sl \bibinfo{series}{Lecture Notes
  in Computer Science}} \bibinfo{volume}{1579}, \bibinfo{publisher}{Springer},
  pp. \bibinfo{pages}{193--207}, \doi{10.1007/3-540-49059-0_14}.

\bibitemdeclare{book}{bradley}
\bibitem{bradley}
\bibinfo{author}{Aaron~R \surnamestart Bradley\surnameend} \&
  \bibinfo{author}{Zohar \surnamestart Manna\surnameend}
  (\bibinfo{year}{2007}): \emph{\bibinfo{title}{The calculus of computation:
  Decision procedures with applications to verification}}.
\newblock \bibinfo{publisher}{Springer}, \doi{10.1007/978-3-540-74113-8}.

\bibitemdeclare{article}{chan1998}
\bibitem{chan1998}
\bibinfo{author}{W.~\surnamestart Chan\surnameend}, \bibinfo{author}{R.~J.
  \surnamestart Anderson\surnameend}, \bibinfo{author}{P.~\surnamestart
  Beame\surnameend}, \bibinfo{author}{S.~\surnamestart Burns\surnameend},
  \bibinfo{author}{F.~\surnamestart Modugno\surnameend},
  \bibinfo{author}{D.~\surnamestart Notkin\surnameend} \&
  \bibinfo{author}{J.~D. \surnamestart Reese\surnameend}
  (\bibinfo{year}{1998}): \emph{\bibinfo{title}{Model checking large software
  specifications}}.
\newblock {\sl \bibinfo{journal}{IEEE Transactions on Software Engineering}}
  \bibinfo{volume}{24}(\bibinfo{number}{7}), pp. \bibinfo{pages}{498--520},
  \doi{10.1109/32.708566}.

\bibitemdeclare{article}{clarke03}
\bibitem{clarke03}
\bibinfo{author}{Edmund~M \surnamestart Clarke\surnameend},
  \bibinfo{author}{Orna \surnamestart Grumberg\surnameend},
  \bibinfo{author}{Somesh \surnamestart Jha\surnameend}, \bibinfo{author}{Yuan
  \surnamestart Lu\surnameend} \& \bibinfo{author}{Helmut \surnamestart
  Veith\surnameend} (\bibinfo{year}{2003}):
  \emph{\bibinfo{title}{Counterexample-guided abstraction refinement for
  symbolic model checking}}.
\newblock {\sl \bibinfo{journal}{Journal of the ACM}}
  \bibinfo{volume}{50}(\bibinfo{number}{5}), pp. \bibinfo{pages}{752--794},
  \doi{10.1145/876638.876643}.

\bibitemdeclare{article}{clarke94}
\bibitem{clarke94}
\bibinfo{author}{Edmund~M \surnamestart Clarke\surnameend},
  \bibinfo{author}{Orna \surnamestart Grumberg\surnameend} \&
  \bibinfo{author}{David~E \surnamestart Long\surnameend}
  (\bibinfo{year}{1994}): \emph{\bibinfo{title}{Model checking and
  abstraction}}.
\newblock {\sl \bibinfo{journal}{ACM Transactions on Programming Languages and
  Systems}} \bibinfo{volume}{16}(\bibinfo{number}{5}), pp.
  \bibinfo{pages}{1512--1542}, \doi{10.1145/186025.186051}.

\bibitemdeclare{book}{clarke99}
\bibitem{clarke99}
\bibinfo{author}{Edmund~M \surnamestart Clarke\surnameend},
  \bibinfo{author}{Orna \surnamestart Grumberg\surnameend} \&
  \bibinfo{author}{Doron \surnamestart Peled\surnameend}
  (\bibinfo{year}{1999}): \emph{\bibinfo{title}{Model checking}}.
\newblock \bibinfo{publisher}{MIT Press}.

\bibitemdeclare{article}{clarke04}
\bibitem{clarke04}
\bibinfo{author}{Edmund~M \surnamestart Clarke\surnameend},
  \bibinfo{author}{Anubhav \surnamestart Gupta\surnameend} \&
  \bibinfo{author}{Ofer \surnamestart Strichman\surnameend}
  (\bibinfo{year}{2004}): \emph{\bibinfo{title}{{SAT}-based
  counterexample-guided abstraction refinement}}.
\newblock {\sl \bibinfo{journal}{IEEE Transactions on Computer-Aided Design of
  Integrated Circuits and Systems}} \bibinfo{volume}{23}(\bibinfo{number}{7}),
  pp. \bibinfo{pages}{1113--1123}, \doi{10.1109/TCAD.2004.829807}.

\bibitemdeclare{incollection}{graf97}
\bibitem{graf97}
\bibinfo{author}{Susanne \surnamestart Graf\surnameend} \&
  \bibinfo{author}{Hassen \surnamestart Saidi\surnameend}
  (\bibinfo{year}{1997}): \emph{\bibinfo{title}{Construction of abstract state
  graphs with {PVS}}}.
\newblock In: {\sl \bibinfo{booktitle}{Computer Aided Verification}}, {\sl
  \bibinfo{series}{Lecture Notes in Computer Science}} \bibinfo{volume}{1254},
  \bibinfo{publisher}{Springer}, pp. \bibinfo{pages}{72--83},
  \doi{10.1007/3-540-63166-6_10}.

\bibitemdeclare{incollection}{hajdu15}
\bibitem{hajdu15}
\bibinfo{author}{\'Akos \surnamestart Hajdu\surnameend},
  \bibinfo{author}{Andr\'as \surnamestart V\"or\"os\surnameend} \&
  \bibinfo{author}{Tam\'as \surnamestart Bartha\surnameend}
  (\bibinfo{year}{2015}): \emph{\bibinfo{title}{New search strategies for the
  {P}etri net {CEGAR} approach}}.
\newblock In: {\sl \bibinfo{booktitle}{Application and Theory of Petri Nets and
  Concurrency}}, {\sl \bibinfo{series}{Lecture Notes in Computer Science}}
  \bibinfo{volume}{9115}, \bibinfo{publisher}{Springer}, pp.
  \bibinfo{pages}{309--328}, \doi{10.1007/978-3-319-19488-2_16}.

\bibitemdeclare{article}{helke2016}
\bibitem{helke2016}
\bibinfo{author}{Steffen \surnamestart Helke\surnameend} \&
  \bibinfo{author}{Florian \surnamestart Kamm{\"u}ller\surnameend}
  (\bibinfo{year}{2016}): \emph{\bibinfo{title}{Verification of statecharts
  using data abstraction}}.
\newblock {\sl \bibinfo{journal}{International Journal of Advanced Computer
  Science and Applications}} \bibinfo{volume}{7}(\bibinfo{number}{1}), pp.
  \bibinfo{pages}{571--583}, \doi{10.14569/IJACSA.2016.070179}.

\bibitemdeclare{article}{latella1999}
\bibitem{latella1999}
\bibinfo{author}{Diego \surnamestart Latella\surnameend},
  \bibinfo{author}{Istvan \surnamestart Majzik\surnameend} \&
  \bibinfo{author}{Mieke \surnamestart Massink\surnameend}
  (\bibinfo{year}{1999}): \emph{\bibinfo{title}{Automatic Verification of a
  Behavioural Subset of UML Statechart Diagrams Using the SPIN Model-checker}}.
\newblock {\sl \bibinfo{journal}{Formal Aspects of Computing}}
  \bibinfo{volume}{11}(\bibinfo{number}{6}), pp. \bibinfo{pages}{637--664},
  \doi{10.1007/s001659970003}.

\bibitemdeclare{inproceedings}{leucker15}
\bibitem{leucker15}
\bibinfo{author}{Martin \surnamestart Leucker\surnameend},
  \bibinfo{author}{Grigory \surnamestart Markin\surnameend} \&
  \bibinfo{author}{MartinR. \surnamestart Neuh\"au{\ss}er\surnameend}
  (\bibinfo{year}{2015}): \emph{\bibinfo{title}{A New Refinement Strategy for
  {CEGAR}-Based Industrial Model Checking}}.
\newblock In: {\sl \bibinfo{booktitle}{Hardware and Software: Verification and
  Testing}}, {\sl \bibinfo{series}{Lecture Notes in Computer Science}}
  \bibinfo{volume}{9434}, \bibinfo{publisher}{Springer}, pp.
  \bibinfo{pages}{155--170}, \doi{10.1007/978-3-319-26287-1_10}.

\bibitemdeclare{incollection}{mcmillan05}
\bibitem{mcmillan05}
\bibinfo{author}{K.L. \surnamestart McMillan\surnameend}
  (\bibinfo{year}{2005}): \emph{\bibinfo{title}{Applications of {C}raig
  Interpolants in Model Checking}}.
\newblock In: {\sl \bibinfo{booktitle}{Tools and Algorithms for the
  Construction and Analysis of Systems}}, {\sl \bibinfo{series}{Lecture Notes
  in Computer Science}} \bibinfo{volume}{3440}, \bibinfo{publisher}{Springer},
  pp. \bibinfo{pages}{1--12}, \doi{10.1007/11494744_2}.

\bibitemdeclare{phdthesis}{meller16}
\bibitem{meller16}
\bibinfo{author}{Yael \surnamestart Meller\surnameend} (\bibinfo{year}{2016}):
  \emph{\bibinfo{title}{Model Checking Techniques for Behavioral UML Models}}.
\newblock Ph.D. thesis, \bibinfo{school}{Israel Institute of Technology}.

\bibitemdeclare{incollection}{meller2014}
\bibitem{meller2014}
\bibinfo{author}{Yael \surnamestart Meller\surnameend}, \bibinfo{author}{Orna
  \surnamestart Grumberg\surnameend} \& \bibinfo{author}{Karen \surnamestart
  Yorav\surnameend} (\bibinfo{year}{2014}): \emph{\bibinfo{title}{Verifying
  Behavioral UML Systems via CEGAR}}.
\newblock In: {\sl \bibinfo{booktitle}{Integrated Formal Methods}},
  \bibinfo{series}{Lecture Notes in Computer Science},
  \bibinfo{publisher}{Springer}, pp. \bibinfo{pages}{139--154},
  \doi{10.1007/978-3-319-10181-1_9}.

\bibitemdeclare{incollection}{demoura08}
\bibitem{demoura08}
\bibinfo{author}{Leonardo \surnamestart de~Moura\surnameend} \&
  \bibinfo{author}{Nikolaj \surnamestart Bj{\o}rner\surnameend}
  (\bibinfo{year}{2008}): \emph{\bibinfo{title}{{Z3}: An Efficient {SMT}
  Solver}}.
\newblock In: {\sl \bibinfo{booktitle}{Tools and Algorithms for the
  Construction and Analysis of Systems}}, {\sl \bibinfo{series}{Lecture Notes
  in Computer Science}} \bibinfo{volume}{4963}, \bibinfo{publisher}{Springer},
  pp. \bibinfo{pages}{337--340}, \doi{10.1007/978-3-540-78800-3_24}.

\bibitemdeclare{article}{nemeth2009}
\bibitem{nemeth2009}
\bibinfo{author}{Erzs{\'e}bet \surnamestart N{\'e}meth\surnameend},
  \bibinfo{author}{Tam{\'a}s \surnamestart Bartha\surnameend},
  \bibinfo{author}{Csaba \surnamestart Fazekas\surnameend} \&
  \bibinfo{author}{Katalin~M. \surnamestart Hangos\surnameend}
  (\bibinfo{year}{2009}): \emph{\bibinfo{title}{Verification of a
  primary-to-secondary leaking safety procedure in a nuclear power plant using
  coloured {P}etri nets}}.
\newblock {\sl \bibinfo{journal}{Reliability Engineering \& System Safety}}
  \bibinfo{volume}{94}(\bibinfo{number}{5}), pp. \bibinfo{pages}{942--953},
  \doi{10.1016/j.ress.2008.10.012}.

\end{thebibliography}
\end{document}